\documentclass[aps,twocolumn,nofootinbib,superscriptaddress]{revtex4-1}
\usepackage{amssymb,amsmath,amstext,amsfonts,color,ulem,textcomp}
\usepackage{graphicx,multirow}

\newcommand{\beq}{\begin{equation}}
\newcommand{\eeq}{\end{equation}}
\newcommand{\bal}{\begin{aligned}}
\newcommand{\eal}{\end{aligned}}
\newcommand{\Lag}{{\mathcal{L}}}

\begin{document}

\title{Destabilization of black holes and stars by generalized Proca fields} 

\author{Sebastian Garcia-Saenz}
\email{s.garcia-saenz@imperial.ac.uk}

\affiliation{Theoretical Physics, Blackett Laboratory, Imperial College London, SW7 2AZ London, U.K.}
\affiliation{Department of Physics, Southern University of Science and Technology, Shenzhen 518055, China}

\author{Aaron Held}
\email{a.held@imperial.ac.uk}

\affiliation{Theoretical Physics, Blackett Laboratory, Imperial College London, SW7 2AZ London, U.K.}

\author{Jun Zhang}
\email{jun.zhang@imperial.ac.uk}

\affiliation{Theoretical Physics, Blackett Laboratory, Imperial College London, SW7 2AZ London, U.K.}
\affiliation{Illinois Center for Advanced Studies of the Universe \&
Department of Physics, University of Illinois at Urbana-Champaign,
Urbana, Illinois 61801, USA}

\begin{abstract}

We demonstrate that black holes and stars in general relativity can be destabilized by perturbations of non-minimally coupled vector fields. Focusing on static and spherically symmetric backgrounds, our analysis shows that black holes with sufficiently small mass and stars with sufficiently high densities are subject to ghost- or gradient-type instabilities. This holds for a large class of Einstein-Proca theories with non-minimal couplings, including generalized Proca models that have sparked attention for their potential role in cosmology and astrophysics. The stability criteria translate into bounds of relevance for low-scale theories of dark energy and for ultra-light dark matter scenarios.

\end{abstract} 

\maketitle

%%%%%%%%%%%%%%%%%%%%%%%%%%%%%%%%%%%%%%%%%%%
%%%%%%%%%%%%%%%%%%%%%%%%%%%%%%%%%%%%%%%%%%%

\textit{Introduction.---}Compact astrophysical objects afford a unique opportunity to probe the existence of new light particles whose couplings to ordinary matter are too weak for direct detection. Observational signatures, including potentially dramatic effects~\cite{Press:1971wr,1971JETPL..14..180Z,Press:1972zz,Damour:1976kh,Zouros:1979iw,Detweiler:1980uk,Yoshino:2012kn,Brito:2015oca}, bear the potential to reveal otherwise hidden gravitationally-induced bosonic condensates. Black holes and compact stars thus constitute precious targets for testing particle physics and cosmology via strong-field gravity and multi-messenger astronomy~\cite{Berti:2015itd,Cardoso:2019rvt}.

Theory needs to inform experiment by mapping out the ``phase diagram'' of black holes and compact stars. 
In a first step, an ambitious research programme drives the development of theoretically consistent bosonic field theories and their interaction with gravity \cite{Deffayet:2013lga,Langlois:2018dxi,Kobayashi:2019hrl}. In a second step, astrophysically relevant solutions are analyzed, asking (i) whether they can account for the observed compact objects and (ii) whether potential instabilities can occur.

Here, we advance the second step and focus on vector bosons~\cite{Galison:1983pa,Holdom:1985ag}. These are predicted by a number of scenarios beyond the Standard Model \cite{Abel:2008ai,Goodsell:2009xc} and serve as viable candidates for dark matter \cite{ArkaniHamed:2008qp,Nelson:2011sf,Arias:2012az} and dark energy \cite{DeFelice:2016yws,DeFelice:2016uil,BeltranJimenez:2017cbn,deFelice:2017paw,Heisenberg:2020xak,Nicosia:2020egv}. If such a light vector particle was to arise from a hidden sector and its dominant interaction with visible matter is mediated by gravity, one is lead to consider Einstein-Proca theory of a massive vector field coupled to gravity. While astrophysical solutions in the minimal version of the theory are constrained by the no-hair theorem \cite{Bekenstein:1971hc,Bekenstein:1972ny}, this does not necessarily apply when including non-minimal couplings, i.e., couplings beyond a covariantization of the kinetic term. 

The general class of Einstein-Proca theories thus accommodates exotic solutions, including hairy black holes \cite{Chagoya:2016aar,Minamitsuji:2016ydr,Babichev:2017rti,Heisenberg:2017xda,Heisenberg:2017hwb,Kase:2018voo,Kase:2018owh}, boson stars \cite{Minamitsuji:2017pdr} and vectorized stars \cite{Nakamura:2017lsf,Kase:2017egk,Kase:2020yhw}. These may be constrained by current and future observations through the beyond-GR effects of Einstein-Proca theories studied, e.g., in~\cite{Pani:2012vp,Witek:2012tr,Brito:2015pxa,Herdeiro:2016tmi,Sanchis-Gual:2017bhw,Cardoso:2018tly,Minamitsuji:2018kof,Sanchis-Gual:2018oui,Santos:2020pmh,CalderonBustillo:2020srq,Caputo:2021efm,Barton:2021wfj}.

Nevertheless, all Einstein-Proca theories which admit the same solutions as General Relativity (GR) (supplemented with a vanishing vector-field background) remain unconstrained. In this situation, the crucial phenomenological question thus concerns the stability of GR solutions.

In this letter, we show that Schwarzschild black holes (and stars) destabilize if their mass (inverse density) drops below a threshold value, related to the non-minimal Einstein-Proca couplings. This allows us to either directly constrain the theory or conclude that GR solutions evolve into non-GR solutions with non-vanishing vector field. 

Although, in the linearized approximation, it is only the vector field and not the metric which suffers from an instability, the backreaction of the vector beyond linear order is expected to render the whole system unstable. Crucially, the uncovered destabilization differs from what is known as ``vectorization"~\cite{Ramazanoglu:2017xbl,Ramazanoglu:2018tig,Annulli:2019fzq,Minamitsuji:2020pak}. In analogy to scalarization \cite{Damour:1993hw}, vectorization occurs due to tachyonic modes, i.e., wrong-sign mass terms. Here, we find that destabilization is always driven by a ghost or gradient mode, i.e., a wrong-sign kinetic or gradient operator. The latter instability is expected to be far more dramatic than the tachyonic one, with potentially unique astrophysical observables. This calls for numerical-relativity investigations (cf.~\cite{East:2017ovw,East:2018glu} for linear Proca fields and \cite{Sanchis-Gual:2015lje,Bosch:2016vcp,Herdeiro:2018wub,Ripley:2019aqj,Ripley:2020vpk,Okounkova:2020rqw,East:2020hgw,Silva:2020omi,Held:2021pht} for related numerical studies in other beyond-GR theories), as one may in principle expect a significant signal in gravitational waves sourced by the exponentially growing vector modes beyond linear order.

This novel destabilization channel and the related astrophysical bounds apply to all Einstein-Proca theories with non-minimal couplings that contribute to the linearized dynamics. This includes Generalized Proca (GP) theory \cite{Tasinato:2014eka,Heisenberg:2014rta,Hull:2015uwa} which has received much attention recently in studies of dark matter and dark energy, as well as on the potential role of new light particles in astrophysical phenomena. We find that destabilization of stellar-mass Schwarzschild black holes constrains cosmological models in which the associated non-minimal coupling is set by the energy scale $\Lambda\sim (M_{\rm Pl}H_0^2)^{1/3}$, where $M_{\rm Pl}$ is the Planck scale and $H_0$ is the Hubble constant. Moreover, if stellar mass black holes acquire transient charges \cite{Wald:1974np,Levin:2018mzg}, destabilization could also constrain fuzzy dark-matter models~\cite{Hu:2000ke}.
\\

%%%%%%%%%%%%%%%%%%%%%%%%%%%%%%%%%%%%%%%%%%%
%%%%%%%%%%%%%%%%%%%%%%%%%%%%%%%%%%%%%%%%%%%

\textit{General quadratic Lagrangian.---}We consider a metric tensor $g_{\mu\nu}$ and vector field $A_{\mu}$ with an action
\begin{align}
\label{eq:full GP action}
S[g,A]&=\int d^4x\sqrt{-g}\bigg[\frac{M_{\rm Pl}^2}{2}\,R-\frac{1}{4}\,F^{\mu\nu}F_{\mu\nu}-\frac{\mu^2}{2}\,A^{\mu}A_{\mu}
\notag\\
&\quad +G_{4,X}A^{\mu}A^{\nu}G_{\mu\nu}-\frac{G_6}{4}\bigg(F^{\mu\nu}F_{\mu\nu}R
\notag\\
&\quad -4F^{\mu\rho}F^{\nu}_{\phantom{\nu}\rho}R_{\mu\nu}+F^{\mu\nu}F^{\rho\sigma}R_{\mu\nu\rho\sigma}\bigg)\bigg] \,.
\end{align}
Here $G_{4,X}$ and $G_6$ are the two constant parameters that define the model, $M_{\rm Pl}=\frac{1}{\sqrt{8\pi G}}$ is the Planck mass and $\mu$ is the mass of the vector field (we assume $\mu^2>0$). For instance, Eq.~\eqref{eq:full GP action} follows from expanding the complete GP theory to quadratic order in the vector field about $\langle A_\mu\rangle=0$ on an arbitrary curved background (see Supplemental Material A).

GP is the complete generalization of the standard Proca theory, i.e.,~its interactions preserve the existence of a (local) frame in which the component $A_0$ is non-dynamical. Although sufficient, this is not necessary for consistency with respect to the number of degrees of freedom \cite{ErrastiDiez:2019trb,ErrastiDiez:2019ttn}, see \cite{Heisenberg:2016eld,Kimura:2016rzw,deRham:2020yet} for alternative extensions. Nevertheless, eq.\ \eqref{eq:full GP action} is the most general vector-tensor model that is (i) quadratic in the vector field, (ii) a function of the vector field and its first derivative only, (iii) at most linear in the undifferentiated curvature.

Condition (i) follows because we are investigating the linear stability of GR solutions without vector condensate. Condition (ii) is a sufficient condition to avoid extra degrees of freedom as in GP theory. Condition (iii) is motivated by our focus on astrophysical GR backgrounds with subleading higher-derivative terms.

Stability and quasi-normal modes of a minimally-coupled Proca field on black-hole spacetimes are studied in~\cite{Konoplya:2005hr, Rosa:2011my, Baumann:2019eav, Percival:2020skc}. The coupling proportional to the Einstein tensor was considered previously e.g.~in \cite{Chagoya:2016aar,Minamitsuji:2016ydr,Babichev:2017rti}, although restricted to unperturbed backgrounds. The interaction terms involving the field strength, which are reminiscent of the Drummond-Hathrell effective action \cite{Drummond:1979pp}, were studied in \cite{Jimenez:2013qsa}. We confirm their results on the stability of GR black holes as a special case of our more general setup.

%%%%%%%%%%%%%%%%%%%%%%%%%%%%%%%%%%%%%%%%%%%
%%%%%%%%%%%%%%%%%%%%%%%%%%%%%%%%%%%%%%%%%%%

\textit{Stability conditions.---}We focus on static and spherically symmetric backgrounds, for which the metric can be chosen as
\beq \label{eq:bkg metric}
g_{\mu\nu}dx^{\mu}dx^{\nu}=-f(r)dt^2+\frac{dr^2}{g(r)}+r^2\left(d\theta^2+\sin^2\theta d\phi^2\right)\,.
\eeq
Perturbations of the metric about GR backgrounds with vanishing vector field decouple and can be ignored. 
The vector field can be decomposed in vector spherical harmonics (see e.g.\ \cite{Nollert:1999ji}),
\beq \label{eq:spherical harmonic decomposition}
A_{\mu}=\sum_{l=0}^{\infty}\sum_{m=-l}^{l}\sum_{I=1}^{4}C^{(I)}_{l,m}(t,r)\left(Z^{(I)}_{l,m}\right)_{\mu}(\theta,\phi)\,.
\eeq
Explicit expressions for $Z^{(I)}_{l,m}$ are given in Supplemental Material B. The mode functions $C^{(I)}_{l,m}$ with $I=1,2,3$ correspond to perturbations with polar parity while $C^{(4)}_{l,m}$ corresponds to an axial-parity mode. Parity is a ``good quantum number''. Hence, polar and axial perturbations decouple at linear order.

The stability of localized perturbations---with physical size much smaller than all the length scales of the background---is dictated by the structure of the causal cones (see \cite{deRham:2017aoj,Babichev:2018uiw,deRham:2019gha, deRham:2020ejn} for related discussions). In other words, to address the question of {\it local} stability one may neglect background variations and evaluate all metric functions at fixed radius $r_0$. The propagator matrix for the mode functions $C^{(I)}_{l,m}$ is defined in Fourier space. Gradient and tachyonic instabilities can be determined by the dispersion relations, defined by the poles of the inverse propagator matrix. The presence of ghosts follows from the matrix of residues. See Supplemental Material C for details. Henceforth, we drop the subscript on the fixed radius $r_0$.

The axial sector has a single degree of freedom. Its dispersion relation follows from the decomposed action,
\beq
\frac{\mathcal{H}_1}{f}\,\omega^2-g\mathcal{H}_2\,k^2-\left(\mathcal{N}_m+\frac{l(l+1)}{r^2}\,\mathcal{N}_j\right)=0 \,,
\eeq
where $\omega$ and $k$ are the comoving (as opposed to proper) frequency and radial wave number, and
\beq\bal \label{eq:H,N defs}
\mathcal{H}_1&=1-G_6\,\frac{g'}{r}\,,\qquad \mathcal{H}_2=1-G_6\,\frac{f'g}{fr} \,, \\
\mathcal{N}_m&=\mu^2+G_{4,X}\left(R-2r^2R^{\theta\theta}\right) \,, \\
\mathcal{N}_j&=1+G_6\left(R-4r^2R^{\theta\theta}+\frac{2(1-g)}{r^2}\right) \,.
\eal\eeq
Here, a prime denotes differentiation with respect to $r$. The curvature terms $R$ and $R^{\theta\theta}$ are known in terms of $f$ and $g$.

For $l\geq1$, only two combinations of the three polar mode functions $C^{(1,2,3)}_{l,m}$ are dynamical. Integrating out the non-dynamical mode, see Supplemental Material B, we can infer the 2-by-2 (inverse) propagator matrix ${\cal P}$. Its components read
\beq\bal
\label{eq:propagatorMatrixPolar}
\mathcal{P}_{11}&= \frac{a_0^2}{g\left({\cal M}_2+{\cal H}_2\frac{l(l+1)}{r^2}\right)}\,\omega^2 \\
&\quad -\frac{fa_0^2}{\left({\cal M}_1+{\cal H}_1\frac{l(l+1)}{r^2}\right)}\,k^2-\sigma_0 \,, \\
\mathcal{P}_{22}&= \frac{{\cal M}_1{\cal H}_1}{fr^2\left({\cal M}_1+{\cal H}_1\frac{l(l+1)}{r^2}\right)}\,\omega^2 \\
&\quad -\frac{g{\cal M}_2{\cal H}_2}{r^2\left({\cal M}_2+{\cal H}_2\frac{l(l+1)}{r^2}\right)}\,k^2-\frac{{\cal N}_m}{r^2} \,, \\
\mathcal{P}_{12}&= \frac{\sigma_0a_0\sqrt{l(l+1)}\left({\cal M}_1{\cal H}_2-{\cal M}_2{\cal H}_1\right)}{r^2\left({\cal M}_1+{\cal H}_1\frac{l(l+1)}{r^2}\right)\left({\cal M}_2+{\cal H}_2\frac{l(l+1)}{r^2}\right)}\,\omega k \,.
\eal\eeq
Here $a_0=\sqrt{\frac{g|{\cal G}_1|}{f}}\,$, $\sigma_0={\rm sign}({\cal G}_1)$ and
\beq\bal \label{eq:G,M defs}
\mathcal{G}_1&=1+2G_6\,\frac{1-g}{r^2} \,,\\
\mathcal{M}_1&=\mu^2-2G_{4,X}\left(\frac{g'}{r}-\frac{1-g}{r^2}\right) \,,\\
\mathcal{M}_2&=\mu^2-2G_{4,X}\left(\frac{f'g}{fr}-\frac{1-g}{r^2}\right) \,.
\eal\eeq
The dispersion relations are defined by the roots $\omega^2_{\pm}$ of the equation ${\rm det}\,{\cal P}=0$.

Monopole perturbations with $l=0$ are special in that only $C^{(1)}_{0,0}$ is present in the polar sector. Its dispersion relation reads
\beq
\frac{|{\cal G}_1|}{f{\cal M}_2}\,\omega^2-\frac{g|{\cal G}_1|}{{\cal M}_1}\,k^2-\sigma_0=0 \,.
\eeq

Notice that the dispersion relations of the polar and axial sector involve the same functions $\mathcal{H}_1$, $\mathcal{H}_2$ and $\mathcal{N}_m$. A priori these functions need not be in any way related---in fact they are not for the monopole modes. This coincidence has important consequences for the stability conditions.

The stability of axial perturbations under ghosts and radial gradients dictates that $\mathcal{H}_1>0$ and $\mathcal{H}_2>0$ for all physical radii. Stability of these modes under angular gradients, which means that $\omega^2$ must be positive in the limit $l\to\infty$, requires $\mathcal{N}_j>0$. Similarly, stability of the polar monopole mode implies that $\mathcal{M}_1>0$ and $\mathcal{M}_2>0$. Given these conditions, it then follows that all polar modes with $l\geq1$ are stable under ghosts and radial gradients, see Supplemental Material C. Further, the stability of these modes under angular gradients gives independent constraints, namely $\mathcal{N}_m>0$ and $\mathcal{G}_1>0$. In turn, these last two conditions imply the absence of tachyonic instabilities for all the perturbations.

An important outcome is that tachyon-like instabilities are absent---for if such modes are excited, they are necessarily accompanied by ghosts and/or gradient-unstable modes with a much faster growth rate. Hence, vector condensates cannot form as a result of a standard vectorization mechanism---which by definition follows from a tachyon- or Jeans-type destabilization---starting from {\it any} static spherically symmetric GR state and for {\it any} Einstein-Proca theory that reduces to eq.\ \eqref{eq:full GP action} at linear order.

%%%%%%%%%%%%%%%%%%%%%%%%%%%%%%%%%%%%%%%%%%%
%%%%%%%%%%%%%%%%%%%%%%%%%%%%%%%%%%%%%%%%%%%

\textit{Black holes.---}For the Schwarzschild black hole (BH) of mass $M$, i.e., for $f=g=1-\frac{r_s}{r}$ with $r_s=2GM$, the stability conditions simplify. Whenever $g=f$ holds, $\mathcal{H}_1=\mathcal{H}_2$ and $\mathcal{M}_1=\mathcal{M}_2$ and the propagator matrix in eq.~\eqref{eq:propagatorMatrixPolar} is diagonal. Moreover, for vacuum GR solutions, the dependence on $G_{4,X}$ drops out; cf.\ eq.\ \eqref{eq:full GP action}. For the Schwarzschild spacetime, one finds that $\mathcal{N}_m=\mathcal{M}_1=\mathcal{M}_2=\mu^2$ are automatically positive, while $\mathcal{H}_1=\mathcal{H}_2=1-\frac{G_6r_s}{r^3}$ and $\mathcal{N}_j=\mathcal{G}_1=1+\frac{2G_6r_s}{r^3}$. Positivity of these functions for all $r\geq r_s$ requires
\beq \label{eq:schw g6 bound}
-\frac{1}{2}<\frac{G_6}{r_s^2}<1 \,,
\eeq
in order for Schwarzschild BHs to be stable. This bound is in agreement with the results of \cite{Jimenez:2013qsa}. It implies that small enough BHs are always unstable whenever $G_6$ is non-zero. 

Order-of-magnitude estimates (assuming validity of the vector theory on all involved scales) reveal that the stability bound in eq.\ \eqref{eq:schw g6 bound} could be of relevance both in late-time cosmology as well as for primordial BHs. In the cosmological setting, non-linear operators are typically controlled by an energy scale $\Lambda\sim (M_{\rm Pl}H_0^2)^{1/3}$, where $H_0$ is the Hubble constant \cite{Nicolis:2004qq,Nicolis:2008in}.\footnote{This estimate is based on scalar-tensor theories and implicitly assumes the existence of a decoupling limit in which the vector-tensor models we consider can be approximated by scalar-tensor interactions.} Taking, for example, $G_6\sim\Lambda^{-2}$, this yields $G_6\sim (10^3\,{\rm km})^2$, implying the destabilization of stellar-mass BHs while supermassive BHs remain stable.
For smaller values of $G_6$, stellar-mass BHs remain stable while primordial BHs in the experimentally preferred range $r_s\sim 10^{-10}\,{\rm m}$ \cite{Carr:2020gox} would still be subject to the instability.
\begin{figure}
  \includegraphics[width=0.4\textwidth]{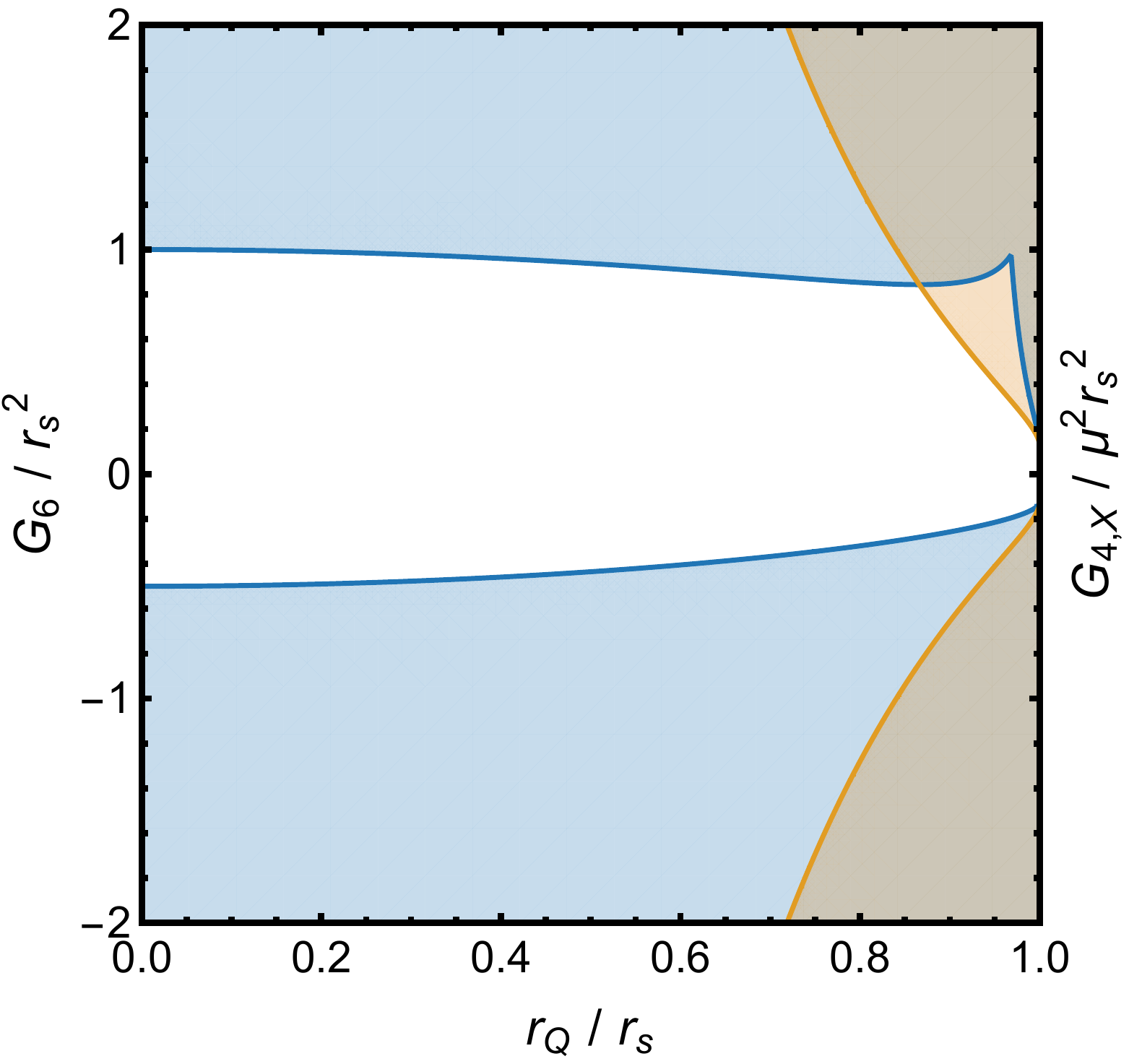}
  \caption{Region plot of the GP parameters for which a destabilization of the RN BH occurs, blue for $G_6$ and orange for $G_{4,X}/\mu^2$ (normalized by the Schwarzschild radius).}
  \label{fig:RNplot}
\end{figure}

Constraining the parameter $G_{4,X}$ requires to look at non-Ricci-flat GR solutions. We consider the Reissner-Nordstr\"om (RN) metric as a first example. Although astrophysical BHs are unlikely to exhibit significant electric (or magnetic) charge, small and transient charges remain viable. For instance, stellar-mass BHs could accrete charges up to of order $10^{-7}$, in units of the BH mass \cite{Levin:2018mzg}, through the Wald mechanism \cite{Wald:1974np} in a merger with a strongly magnetized neutron star.\footnote{In \cite{Wald:1974np,Levin:2018mzg}, the charging effect requires a spinning BH, however, a more recent study \cite{Chen:2021sya} has shown that rotation is not needed and that the relative motion between the coalescing BH and neutron star can generate charges of comparable magnitude.}

The RN metric is defined by $f=g=1-\frac{r_s}{r}+\frac{r_Q^2}{4r^2}$. Here, $r_Q=2\sqrt{G}\,Q$ in terms of the hole's electric charge and we recall the extremality bound $r_Q\leq r_s$. The stability conditions now depend on the scale $r_Q$, cf.\ fig.\ \ref{fig:RNplot}, and Supplemental Material D for the analytic expressions. For $G_6$, we observe a non-trivial dependence of the stability bounds on the charge, cf.~\cite{Jimenez:2013qsa}; in particular, they are most restrictive for an extremal BH, for which $|G_6|/r_s^2<1/8$.

More interestingly, we find a novel bound on $G_{4,X}$,
\beq
\label{eq:charged-BH-bound}
\frac{|G_{4,X}|}{\mu^2 r_s^2} < \frac{\left(1+\sqrt{1-(r_Q/r_s)^2}\right)^4}{8(r_Q/r_s)^2} \,.
\eeq
Remarkably, for any fixed $G_{4,X}$, $r_s$, and $r_Q$, this bound implies a lower limit on the vector-boson mass $\mu$. As a concrete example, for $G_{4,X}={\cal O}(1)$ , as typically considered in the literature, stability of a stellar-mass BH with $r_s\sim 10\,{\rm km}$ that acquires the aforementioned estimate $r_Q\sim 10^{-7}\,r_s$ implies $\mu\sim10^{-17}\,{\rm eV}$ as the critical vector-boson mass. Comparison with the typical mass range $10^{-22}-10^{-20}\,{\rm eV}$ for fuzzy dark matter \cite{Hu:2000ke} exemplifies the significance of eq.~\eqref{eq:charged-BH-bound} for the study of ultra-light particles. We note that $G_{4,X}$ may not be independent of $\mu$: if the operators that break gauge invariance were to arise from a Higgs-type mechanism, we would expect $G_{4,X}\propto \mu^2$ \cite{deRham:2018qqo} and our stability criteria would not directly constrain the mass $\mu$ but rather the scale of symmetry breaking.

%%%%%%%%%%%%%%%%%%%%%%%%%%%%%%%%%%%%%%%%%%%
%%%%%%%%%%%%%%%%%%%%%%%%%%%%%%%%%%%%%%%%%%%

\textit{Stars.---}We have analyzed the stability conditions for static perfect fluid stars governed by the TOV equations. Although for generic equations of state (EoS)---relating the pressure $p$ to the density $\rho$---the metric cannot be determined in analytic form, critical values for the parameters $G_6$ and $G_{4,X}$ can still be obtained if one assumes that the functions that determine the stability are minimized at the center of the star. We have checked analytically that this assumption is correct for a uniform-density star, and also numerically for a polytropic star with EoS $p=K\rho^{5/3}$, see Supplemental Material D. It is plausible that the assumption is true for all realistic EoS, including ones for imperfect fluids, and we plan to come back to this question in a dedicated work.

\begin{figure}
  \includegraphics[width=0.45\textwidth]{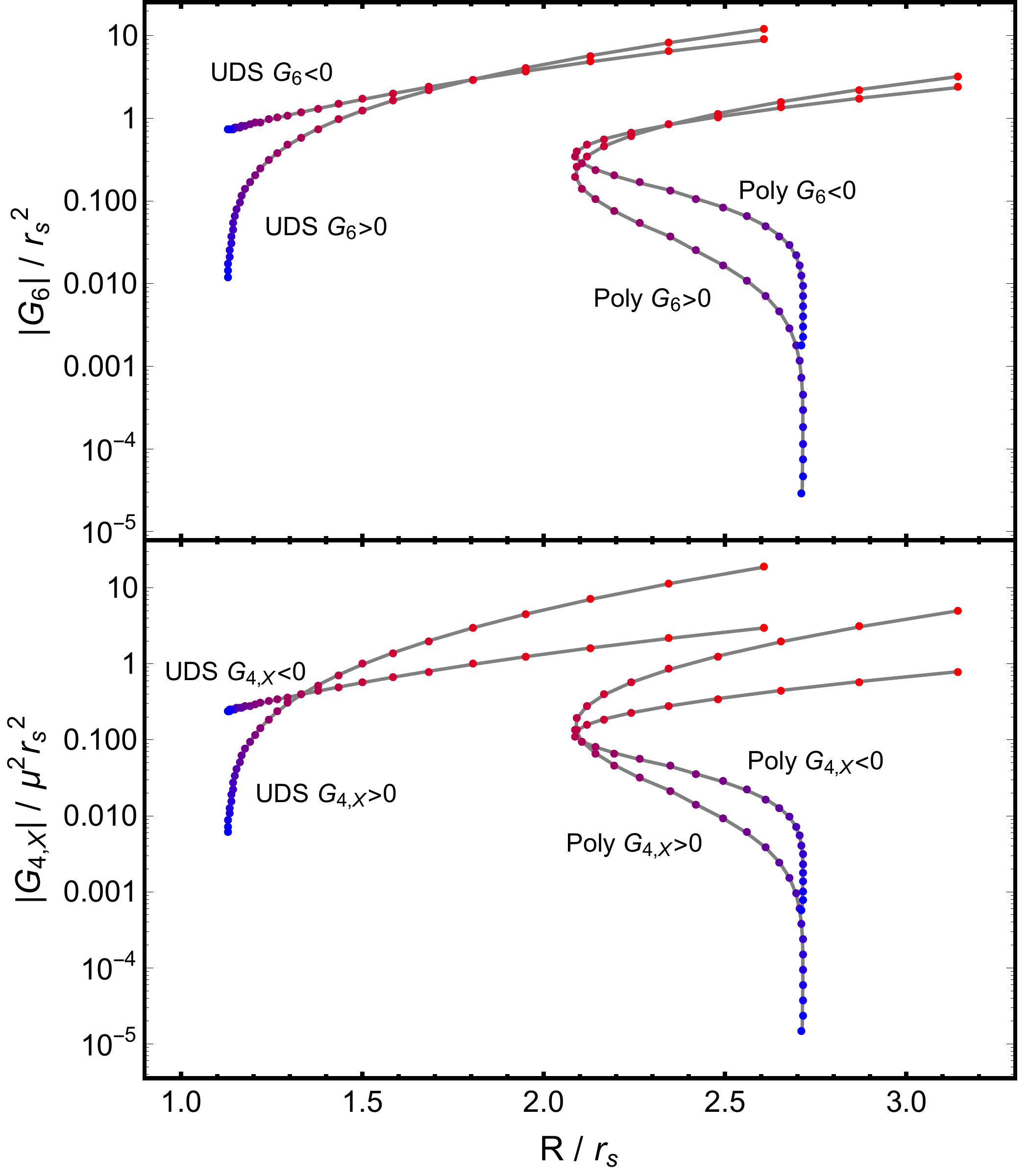}
  \caption{Critical values of the GP parameters for which an instability is triggered in stars modeled by uniform density (``UDS'') and $\gamma=5/3$ polytropic index (``Poly'') as inferred from eq.\ \eqref{eq:stars bounds}. Colored points label different values of the central pressure, ranging from $10^{-2}$ (red; upper-right end) to $10^4$ (blue; lower end) in arbitrary units such that $K=1$ (the constant appearing in the polytropic EoS). Despite this arbitrariness, the comparison between different pressures and between the two stellar models is meaningful.}
  \label{fig:starsplot}
\end{figure}

Under this premise, we can infer the following bounds on the GP coupling constants:
\beq\begin{gathered} \label{eq:stars bounds}
-\frac{3}{2\rho_c}<\frac{G_6}{M_{\rm Pl}^2}<\frac{3}{\rho_c+3p_c} \,, \\
-\frac{1}{2\rho_c}<\frac{G_{4,X}}{\mu^2 M_{\rm Pl}^2}<\frac{1}{2p_c} \,,
\end{gathered}\eeq
where $p_c$ and $\rho_c$ are the pressure and density at the center. Fig.\ \ref{fig:starsplot} shows the critical values of $G_6$ and $G_{4,X}/\mu^2$ for stellar models with uniform-density and $\gamma=5/3$ polytropic EoS, plotted as functions of the normalized star's radius. We observe an interesting dependence on the EoS, with the bounds for a polytrope being up to three orders of magnitude stronger than for a uniform-density star with the same central pressure and density.

The stability window for both coupling parameters shrinks to zero as the star's central pressure and density increase. For a neutron star with $\rho_c\sim 10^{18}\,{\rm kg\,m^{-3}}\sim 10^{-76}\,M_{\rm Pl}^4$ one has $\Lambda/M_{\rm Pl} \gtrsim 10^{-38}$ if we take $|G_6|\sim |G_{4,X}|/\mu^2 \sim \Lambda^{-2}$. This bound on $\Lambda$ may seem mild but again could be violated in very low-scale models like the ones envisioned in cosmology and in the context of ultra-light dark matter.

%%%%%%%%%%%%%%%%%%%%%%%%%%%%%%%%%%%%%%%%%%%
%%%%%%%%%%%%%%%%%%%%%%%%%%%%%%%%%%%%%%%%%%%

\textit{Discussion.---}We identify a novel destabilization channel for static and spherically-symmetric GR backgrounds triggered by non-minimally coupled vector perturbations. Any non-vanishing non-minimal coupling destabilizes small enough BHs and dense enough stars. The implied astrophysical constraints ultimately depend on the scales at hand. We find relevant constraints for theories of dark energy and ultra-light dark matter. The non-minimal couplings that source destabilization, cf.\ eq.~\eqref{eq:full GP action}, naturally appear in all of these scenarios, unless the model is fine-tuned to avoid them.

More specifically, avoiding instabilities of stellar-mass BHs and/or neutron stars constrains the respective non-minimal coupling at cosmologically relevant scales set by $\Lambda\sim (M_{\rm Pl}H_0^2)^{1/3}$, with $M_{\rm Pl}$ the Planck scale and $H_0$ the Hubble constant. In turn, transient charges, potentially induced by nearby strongly magnetized neutron stars, imply further destabilization constraints involving the Proca mass and are of relevance for ultra-light vector dark-matter models. 

Notably, destabilization differs from vectorization. The latter describes a transition between GR and non-GR solutions via a tachyonic growth mode. Here, we find that a potential tachyonic instability is always accompanied by a dominant ghost or gradient instability. Hence, destabilization is controlled by the highest growth rates in the problem. The timescale and fate of the instability thus remain uncertain. 

We emphasize that the destabilization channel concerns only GR solutions and does not constrain solutions with non-trivial vector hair.
Formally, we have only shown a linear instability in the vector field, and not in the metric. However, as one may expect for ghost/gradient modes, interactions beyond linear order will generically destabilize the full system. Future non-linear studies are necessary to strictly discard the possibility that the vector field may settle into a condensed state with the GR metric being kept intact, as it occurs with so-called ``stealth'' solutions \cite{Babichev:2013cya,Chagoya:2017ojn,Minamitsuji:2018vuw,Minamitsuji:2020hpl}.

From the perspective of radiative corrections, the inclusion of higher-derivative operators may quench the instability, similarly to the phenomenon of ghost condensation \cite{ArkaniHamed:2003uy}. This possibility calls for a detailed study to determine the role of higher-order operators. Assuming the transition can be made sense of in a controlled theoretical framework, our analysis makes a strong case for simulating Einstein-Proca theories in numerical relativity.

There are also several avenues for future work within the present set-up of linear perturbations about GR backgrounds. This includes (i) studying a broader set of stellar models in order to verify the robustness of our bounds in eq.\ \eqref{eq:stars bounds}; (ii) effects of a cosmological constant, potentially related to extended vector fields in holographic models (see e.g.\ \cite{Jing:2010zp,Jing:2010cx,Zhao:2012cn}); and, of course, (iii) an extension to non-static systems, in particular rotating BHs and stars.
\\

%%%%%%%%%%%%%%%%%%%%%%%%%%%%%%%%%%%%%%%%%%%
%%%%%%%%%%%%%%%%%%%%%%%%%%%%%%%%%%%%%%%%%%%

\textit{Acknowledgments.---}We would like to thank Claudia de Rham, Lavinia Heisenberg, Shinji Mukohyama and an anonymous referee for useful comments. SGS and JZ are supported by the European Union's Horizon 2020 Research Council grant 724659 MassiveCosmo ERC-2016-COG. AH is supported by a Royal Society Newton International fellowship [NIF{\textbackslash}R1{\textbackslash}191008].

%%%%%%%%%%%%%%%%%%%%%%%%%%%%%%%%%%%%%%%%%%%
%%%%%%%%%%%%%%%%%%%%%%%%%%%%%%%%%%%%%%%%%%%

\appendix

\section{Generalized Proca Lagrangian} \label{sec:appA}

The Lagrangian of Generalized Proca (GP) theory in four dimensions is defined as \cite{Tasinato:2014eka,Heisenberg:2014rta}
\beq\bal \label{eq:app full GP action}
S[g,A]&=\int d^4x\sqrt{-g}\bigg[\frac{M_{\rm Pl}^2}{2}\,R-\frac{1}{4}\,F^{\mu\nu}F_{\mu\nu} \\
&\quad -\frac{\mu^2}{2}\,A^{\mu}A_{\mu}+\sum_{I=2}^6 \Lag_I[g,A]\bigg] \,,
\eal\eeq
where
\beq\bal \label{eq:app GP terms def}
\Lag_2&=G_2(X,{\cal F},{\cal G})\,, \\
\Lag_3&=G_3(X)\nabla_{\mu}A^{\mu}\,, \\
\Lag_4&=G_4(X)R+G_{4,X}(X)\Big[(\nabla_{\mu}A^{\mu})^2-\nabla_{\mu}A^{\nu}\nabla_{\nu}A^{\mu}\Big]\,, \\
\Lag_5&=G_5(X)G^{\mu\nu}\nabla_{\mu}A_{\nu}-\frac{G_{5,X}(X)}{6}\Big[(\nabla_{\mu}A^{\mu})^3 \\
&\quad -3\nabla_{\rho}A^{\rho}\nabla_{\mu}A^{\nu}\nabla_{\nu}A^{\mu}+2\nabla_{\mu}A^{\nu}\nabla_{\nu}A^{\rho}\nabla_{\rho}A^{\mu}\Big]\,, \\
\Lag_6&=G_6(X)\widetilde{R}^{\mu\nu\rho\sigma}\nabla_{\mu}A_{\nu}\nabla_{\rho}A_{\sigma} \\
&\quad +\frac{G_{6,X}(X)}{2}\,\widetilde{F}^{\mu\nu}\widetilde{F}^{\rho\sigma}\nabla_{\mu}A_{\rho}\nabla_{\nu}A_{\sigma} \,, \\
\eal\eeq
with the definitions
\beq\begin{gathered}
X:=-\frac{1}{2}\,A^{\mu}A_{\mu} \,,\qquad {\cal F}:=-\frac{1}{4}\,F^{\mu\nu}F_{\mu\nu} \,, \\
{\cal G}:=A^{\mu}A^{\nu}F_{\mu}^{\phantom{\mu}\rho}F_{\nu\rho} \,,\qquad \widetilde{F}^{\mu\nu}:=\frac{1}{2}\,\epsilon^{\mu\nu\mu'\nu'}F_{\mu'\nu'} \\
\widetilde{R}^{\mu\nu\rho\sigma}:=\frac{1}{4}\,\epsilon^{\mu\nu\mu'\nu'}\epsilon^{\rho\sigma\rho'\sigma'}R_{\mu'\nu'\rho'\sigma'} \,.
\end{gathered}\eeq
The expression $G_{4,X}(X)$ means the derivative of the function $G_4(X)$ with respect to its argument, and similarly for $G_{5,X}$ and $G_{6,X}$. Note that in \eqref{eq:app full GP action} we have written explicitly the Einstein-Hilbert and Proca Lagrangians so that we may take $G_4(0)=G_{2,X}(0,0,0)=G_{2,{\cal F}}(0,0,0)=0$. Observe also that $G_3(0)$ and $G_5(0)$ multiply total derivatives and may be ignored in a perturbative expansion in powers of the vector field.

Expanding the full action to quadratic order in $A_{\mu}$ and its derivative gives eq.\ \eqref{eq:full GP action} in the main text after one notices that
\beq
(\nabla_{\mu}A^{\mu})^2-\nabla_{\mu}A^{\nu}\nabla_{\nu}A^{\mu}=R_{\mu\nu}A^{\mu}A^{\nu}+{\rm t.d.} \,,
\eeq
(``t.d.'' means total derivative) and
\beq\bal
\,& \frac{1}{4}\,\epsilon^{\mu\nu\mu'\nu'}\epsilon^{\rho\sigma\rho'\sigma'}F_{\mu\nu}F_{\rho\sigma}R_{\mu'\nu'\rho'\sigma'} \\
&=-F^{\mu\nu}F_{\mu\nu}R+4F^{\mu\rho}F^{\nu}_{\phantom{\nu}\rho}R_{\mu\nu}-F^{\mu\nu}F^{\rho\sigma}R_{\mu\nu\rho\sigma} \,.
\eal\eeq

The conclusion is that the GP Lagrangian reduces, when truncated to quadratic order in the vector field, to the standard Proca theory plus a sum of non-minimal couplings involving $A_{\mu}$, $F_{\mu\nu}$ and the curvature tensor. These non-minimal coupling terms must appear in specific combinations---proportional to the Einstein tensor and the dual Riemann tensor---in order to avoid additional degrees of freedom.

We mentioned in the main text that other extensions of the linear Proca theory will not give additional operators within the framework specified by our assumptions. This is clear for the models proposed in \cite{Heisenberg:2016eld,Kimura:2016rzw} which modify GP theory with terms that do not contribute at quadratic order. On the other hand, the model of \cite{deRham:2020yet} is genuinely independent from GP (in the sense that the Lagrangians cannot be matched by any choice of parameters); nevertheless, when expanded to quadratic order the two models are not inequivalent and hence the proposal of \cite{deRham:2020yet} does fall within our class once complemented with appropriate non-minimal curvature couplings.

%%%%%%%%%%%%%%%%%%%%%%%%%%%%%%%%%%%%%%%%%%%
%%%%%%%%%%%%%%%%%%%%%%%%%%%%%%%%%%%%%%%%%%%

\section{Expansion of the action in spherical harmonics} \label{sec:appB}

\subsection{Vector spherical harmonics}

The decomposition of the Proca field in vector spherical harmonics is given in eq.\ \eqref{eq:spherical harmonic decomposition} in the main text. In our conventions the vector harmonic functions are given by
\beq\bal
(Z^{(1)}_{l,m})_{\mu}&=\delta^t_{\mu}Y_{l,m}(\theta,\phi) \,, \\
(Z^{(2)}_{l,m})_{\mu}&=\delta^r_{\mu}Y_{l,m}(\theta,\phi) \,, \\
(Z^{(3)}_{l,m})_{\mu}&=\frac{1}{\sqrt{l(l+1)}}\,\partial_{\mu}Y_{l,m}(\theta,\phi) \,, \\
(Z^{(4)}_{l,m})_{\mu}&=\frac{1}{\sqrt{l(l+1)}}\big[-\csc\theta\, \delta^{\theta}_{\mu}\partial_{\phi}Y_{l,m}(\theta,\phi) \\
&\quad +\sin\theta\, \delta^{\phi}_{\mu}\partial_{\theta}Y_{l,m}(\theta,\phi)\big] \,,
\eal\eeq
with $Y_{l,m}$ denoting the standard spherical harmonic functions which solve the Laplace equation on the sphere,
\beq
\frac{1}{\sin\theta}\frac{\partial}{\partial\theta}\left(\sin\theta\,\frac{\partial Y_{l,m}}{\partial\theta}\right)+\frac{1}{\sin^2\theta}\,\frac{\partial^2 Y_{l,m}}{\partial\phi^2}+l(l+1)Y_{l,m}=0 \,.
\eeq
The three functions $Z^{(1,2,3)}_{l,m}$ have polar or even parity, i.e.\ they acquire a factor $(-1)^l$ under space inversions $(\theta,\phi)\to(\pi-\theta,\pi+\phi)$, while $Z^{(4)}_{l,m}$ has axial or odd parity, acquiring a factor $(-1)^{l+1}$ under inversions.

We recall the basic orthogonality property of the spherical harmonics,
\beq
\int d\Omega\, Y^{*}_{l,m}Y_{l',m'}=\delta_{l,l'}\delta_{m,m'} \,,
\eeq
and also that $Y^{*}_{l,m}=(-1)^mY_{l,-m}$. It then follows that the 4-vector spherical harmonics, with the normalization given above, satisfy
\beq
\int d\Omega\, (Z^{(I)}_{l,m})^{*}_{\mu}M_Z^{\mu\nu}(Z^{(J)}_{l',m'})_{\mu}=\delta_{l,l'}\delta_{m,m'}\delta^{I,J}\,,
\eeq
the inner product being defined by the matrix $M_Z^{\mu\nu}={\rm diag}\left(1,1,1,\csc^2\theta\right)^{\mu\nu}$. The vector spherical harmonics also inherit the conjugation property,
\beq
(Z^{(I)}_{l,m})^{*}_{\mu}=(-1)^m(Z^{(I)}_{l,-m})_{\mu} \,.
\eeq
The reality of the field $A_{\mu}$ then implies that $C^{(I)}_{l,-m}=(-1)^mC^{(I)*}_{l,m}$.

\subsection{Axial perturbations}

Expanding the complete Lagrangian in terms of the mode functions we obtain the following result for the axial sector:
\beq\bal
S_{\rm axi}&=\frac{1}{2}\int dt dr\sqrt{\frac{f}{g}}\sum_{l,m}(-1)^m\bigg[\frac{\mathcal{H}_1}{f}\,|\dot{C}^{(4)}_{l,m}|^2 \\
&\quad -g\mathcal{H}_2\,|C^{(4)\prime}_{l,m}|^2 -\left(\mathcal{N}_m+\frac{l(l+1)}{r^2}\,\mathcal{N}_j\right)|C^{(4)}_{l,m}|^2\bigg] \,,
\eal\eeq
with the functions $\mathcal{H}_1$, $\mathcal{H}_2$, $\mathcal{N}_m$ and $\mathcal{N}_j$ as defined in the main text.

\subsection{Polar perturbations}

The Lagrangian for the polar perturbations $C^{(1)}_{l,m}$, $C^{(2)}_{l,m}$ and $C^{(3)}_{l,m}$ is given by
\beq\bal
S_{\rm pol}&=\frac{1}{2}\int dt dr\,r^2\sqrt{\frac{f}{g}}\sum_{l,m}(-1)^m\bigg[\frac{g}{f}\,{\cal G}_1\left|\dot{C}^{(2)}_{l,m}-C^{(1)\prime}_{l,m}\right|^2 \\
&\quad +\frac{1}{fr^2}\,{\cal H}_1\left|\dot{C}^{(3)}_{l,m}-\sqrt{l(l+1)}\,C^{(1)}_{l,m}\right|^2 \\
&\quad -\frac{g}{r^2}\,{\cal H}_2\left|C^{(3)\prime}_{l,m}-\sqrt{l(l+1)}\,C^{(2)}_{l,m}\right|^2 \\
&\quad +\frac{1}{f}\,{\cal M}_1|C^{(1)}_{l,m}|^2-g\,{\cal M}_2|C^{(2)}_{l,m}|^2-\frac{{\cal N}_m}{r^2}\,|C^{(3)}_{l,m}|^2\bigg] \,,
\eal\eeq
and the functions $\mathcal{M}_1$, $\mathcal{M}_2$ and $\mathcal{G}_1$ can be found in the Letter.

This Lagrangian is degenerate in the sense that not all among the three mode functions are dynamical and, as explained in the text, it is useful to integrate out the non-dynamical mode. The monopole sector with $l=0$ is preculiar because $C^{(3)}_{0,0}\equiv 0$, so we start by treating this case separately. The trick is to introduce the auxiliary field
\beq \label{eq:B0 def}
B_{0,0}:=a_0\left(\dot{C}^{(2)}_{0,0}-C^{(1)\prime}_{0,0}\right) \,,
\eeq
with $a_0:=\sqrt{\frac{g|{\cal G}_1|}{f}}\,$. This can then be incorporated in the Lagrangian as
\beq\bal
S_{\rm pol}^{(l=0)}&=\frac{1}{2}\int dt dr\,r^2\sqrt{\frac{f}{g}}\bigg\{-\sigma_0|B_{0,0}|^2 \\
&\quad +\sigma_0a_0\left[B_{0,0}^{*}\left(\dot{C}^{(2)}_{0,0}-C^{(1)\prime}_{0,0}\right)+{\rm c.c.}\right] \\
&\quad +\frac{1}{f}\,{\cal M}_1|C^{(1)}_{0,0}|^2-g{\cal M}_2|C^{(2)}_{0,0}|^2\bigg\} \,,
\eal\eeq
with $\sigma_0:={\rm sign}({\cal G}_1)$. Variation with respect to $B_{0,0}^{*}$ gives \eqref{eq:B0 def}, which may be substituted back to recover the original action, proving that the two are indeed equivalent. Alternatively, from the latter form of the action we can integrate out $C^{(1)}_{0,0}$ and $C^{(2)}_{0,0}$ since now their eqs.\ of motion are algebraic:
\beq\bal
C^{(1)}_{0,0}&=-\frac{\sigma_0}{r^2\sqrt{f/g}}\,\frac{f}{{\cal M}_1}\left(r^2\sqrt{f/g}\,a_0B_{0,0}\right)' \,, \\
C^{(2)}_{0,0}&=-\sigma_0\,\frac{a_0}{g{\cal M}_2}\,\dot{B}_{0,0} \,,
\eal\eeq
and we obtain
\beq\bal
S_{\rm pol}^{(l=0)}&=\frac{1}{2}\int dt dr\,r^2\sqrt{\frac{f}{g}}\Bigg[\frac{|{\cal G}_1|}{f{\cal M}_2}\,|\dot{B}_{0,0}|^2-\sigma_0|B_{0,0}|^2 \\
&\quad -\frac{g|{\cal G}_1|}{{\cal M}_1}\left|B_{0,0}'+\frac{(r^2\sqrt{f/g}\,a_0)'}{r^2\sqrt{f/g}\,a_0}\,B_{0,0}\right|^2 \Bigg] \,,
\eal\eeq
for the Lagrangian describing the dynamics of the monopole polar mode.

For generic higher multipoles we can carry out the same procedure in order to remove the non-dynamical mode. We define
\beq \label{eq:B def}
B_{l,m}:=a_0\left(\dot{C}^{(2)}_{l,m}-C^{(1)\prime}_{l,m}\right)\,,\quad C_{l,m}:=C^{(3)}_{l,m}\,,
\eeq
and solving for $C^{(1)}_{l,m}$ and $C^{(2)}_{l,m}$ from their eqs.\ of motion now yields
\begin{widetext}
\beq\bal
C^{(1)}_{l,m}&=\frac{f}{\left({\cal M}_1+{\cal H}_1\frac{l(l+1)}{r^2}\right)}\bigg[-\frac{\sigma_0}{r^2\sqrt{f/g}}\left(r^2\sqrt{f/g}\,a_0B_{l,m}\right)' +\frac{{\cal H}_1\sqrt{l(l+1)}}{fr^2}\,\dot{C}_{l,m}\bigg] \,, \\
C^{(2)}_{l,m}&=\frac{1}{g\left({\cal M}_2+{\cal H}_2\frac{l(l+1)}{r^2}\right)}\bigg[-\sigma_0a_0\dot{B}_{l,m} +\frac{g{\cal H}_2\sqrt{l(l+1)}}{r^2}\,{C}'_{l,m}\bigg] \,.
\eal\eeq
Substituting back in the action we eventually find
\beq\bal \label{eq:app polar lagrangian}
S_{\rm pol}^{(l>0)}&=\frac{1}{2}\int dt dr\,r^2\sqrt{\frac{f}{g}}\sum_{l,m}(-1)^m\Bigg[\frac{a_0^2}{g\left({\cal M}_2+{\cal H}_2\frac{l(l+1)}{r^2}\right)}\,|\dot{B}_{l,m}|^2 -\frac{fa_0^2}{\left({\cal M}_1+{\cal H}_1\frac{l(l+1)}{r^2}\right)}\left|B_{l,m}'+\frac{(r^2\sqrt{f/g}\,a_0)'}{r^2\sqrt{f/g}\,a_0}\,B_{l,m}\right|^2 \\
&\quad +\frac{{\cal M}_1{\cal H}_1}{fr^2\left({\cal M}_1+{\cal H}_1\frac{l(l+1)}{r^2}\right)}\,|\dot{C}_{l,m}|^2 -\frac{g{\cal M}_2{\cal H}_2}{r^2\left({\cal M}_2+{\cal H}_2\frac{l(l+1)}{r^2}\right)}\,|{C}'_{l,m}|^2-\sigma_0|B_{l,m}|^2-\frac{{\cal N}_m}{r^2}\,|C_{l,m}|^2 \\
&\quad -\frac{\sigma_0a_0{\cal H}_2\sqrt{l(l+1)}}{r^2\left({\cal M}_2+{\cal H}_2\frac{l(l+1)}{r^2}\right)}\,\left(\dot{B}^{*}_{l,m}C'_{l,m}+{\rm c.c.}\right) +\frac{\sigma_0{\cal H}_1\sqrt{l(l+1)}}{r^4\sqrt{f/g}\left({\cal M}_1+{\cal H}_1\frac{l(l+1)}{r^2}\right)}\,\left((r^2\sqrt{f/g}\,a_0{B}^{*}_{l,m})'\dot{C}_{l,m}+{\rm c.c.}\right) \Bigg] \,,
\eal\eeq
\end{widetext}
which as claimed contains two dynamical modes for each $l,m$.

%%%%%%%%%%%%%%%%%%%%%%%%%%%%%%%%%%%%%%%%%%%
%%%%%%%%%%%%%%%%%%%%%%%%%%%%%%%%%%%%%%%%%%%

\section{Stability conditions and matrix of propagators} \label{sec:appC}

Consider a fully generic two-derivative quadratic Lagrangian,
\beq
\Lag=-\frac{1}{2}\,{\cal G}^{\mu\nu}_{\phantom{\mu\nu}IJ}\partial_{\mu}\phi^I\partial_{\nu}\phi^J \,.
\eeq
The fields $\phi^I$ are not necessarily scalars, i.e.\ they may be components of a set of tensor fields. The coordinates $x^{\mu}$ are not necessarily Cartesian, although we are primarily interested in the situation where $\partial/\partial x^0$ is timelike and $\partial/\partial x^i$ is spacelike. In principle the tensor ${\cal G}^{\mu\nu}_{\phantom{\mu\nu}IJ}$ may be a function of the coordinates, but for the purpose of determining the presence of ghost and gradient-type instabilities it suffices to assume it is a constant as we explained in the main text.

The inverse Fourier-space propagator is
\beq
(\Delta^{-1})_{IJ}={\cal G}^{\mu\nu}_{\phantom{\mu\nu}IJ}k_{\mu}k_{\nu} \,.
\eeq
Inverting gives the propagator $\Delta^{IJ}$, more precisely the matrix of propagators. The poles of $\Delta^{IJ}$ correspond to the physical particles. By ``poles'' of a matrix we mean the values of $\omega^2$ for which the inverse determinant vanishes. Thus the dispersion relations, which determine the particle spectrum and the causal cone structure, are given by the solutions $\omega^2(k^2)$ of
\beq
{\rm det}\,\Delta^{-1}=0 \,.
\eeq

Gradient instabilities can be determined unambiguously from the dispersion relations. Ghost instabilities, on the other hand, are ambiguous in that they make reference to the orientation of the causal cones relative to another reference particle sector, which is by assumption ``healthy''. If we take this reference sector to be an ordinary scalar field (but any garden-variety field would do),
\beq
\Lag_{\rm ref}=-\frac{1}{2}\,\eta^{\mu\nu}\partial_{\mu}\chi\partial_{\nu}\chi \,,
\eeq
its propagator, as defined above, is obviously $\Delta_{\chi}=\frac{1}{-\omega^2+|\vec{k}|^2}\,$. The dispersion relation is given by the pole, $\omega^2=|\vec{k}|^2$, and there is of course no gradient instability.

Upon quantization, the norms of the physical modes are inferred from the residues of the propagator at the particles' poles. The signs of the norms are conventional---only the {\it relative} signs are important. We choose to define our reference field to have unit norm, and therefore the residue matrix from which the norms are inferred must be given by
\beq
(R_\alpha)^{IJ}=-\lim_{\omega^2\to \omega^2_{\alpha}}(\omega^2-\omega^2_{\alpha})\Delta^{IJ} \,,
\eeq
where $\omega^2_{\alpha}$ is the $\alpha$-th pole of the propagator.

For our reference field $\chi$ we then clearly have $R=1$, as desired. Because this field is by definition ``healthy'', any other dynamical mode having a negative residue (more precisely, a residue matrix with one or more negative eigenvalues) is by definition a ghost. 

Let us apply this to the following 2D toy model:
\beq \label{eq:app toy lagrangian}
\Lag=\frac{1}{2}\,\dot{b}^2+\frac{1}{2}\,\dot{c}^2-\frac{\alpha}{2}\,b^{\prime2}-\frac{\beta}{2}\,c^{\prime2}+\frac{\gamma}{2}\left(b'\dot{c}+\dot{b}c'\right) \,.
\eeq
For generic parameters, this Lagrangian cannot be diagonalized via a local field redefinition (in particular, such a redefinition does not exist whenever $\alpha\neq\beta$). But as we have explained, the particle spectrum and its stability can be determined from the propagator alone. Comparing \eqref{eq:app toy lagrangian} and \eqref{eq:app polar lagrangian} we see that this actually serves as a proxy model for the polar Lagrangian that we sought to analyze.

In the field basis $\phi^I=(b,c)$ we have
\beq
(\Delta^{-1})_{IJ}=\left(\begin{array}{cc}
-\omega^2+\alpha k^2 & 2\gamma\omega k \\
2\gamma\omega k & -\omega^2+\beta k^2 \end{array}\right) \,,
\eeq
so that
\beq
\Delta^{IJ}=-\frac{1}{\cal D}\left(\begin{array}{cc}
\omega^2-\beta k^2 & 2\gamma\omega k \\
2\gamma\omega k & \omega^2-\alpha k^2 \end{array}\right) \,,
\eeq
with determinant
\beq\bal
{\cal D}&=(\omega^2-\alpha k^2)(\omega^2-\beta k^2)-4\gamma^2\omega^2k^2 \\
&=(\omega^2-\omega^2_{+})(\omega^2-\omega^2_{-}) \,,
\eal\eeq
and for the roots we find
\beq
\frac{\omega_{\pm}^2}{k^2}=\frac{1}{2}\left[\alpha+\beta+4\gamma^2\pm\sqrt{(\alpha+\beta+4\gamma^2)^2-4\alpha\beta}\right] \,.
\eeq
Absence of gradient-unstable modes means that these solutions must be positive (a solution with $\omega^2=0$ would signal a degeneracy; we ignore this possibility as it would require a separate analysis). This restricts the parameters by the inequalities
\beq \label{eq:app grad instability}
\alpha\beta>0\,,\qquad \alpha+\beta+4\gamma^2>2\sqrt{\alpha\beta}\,.
\eeq

Next we define the residue matrices,
\beq\bal
(R_{\pm})^{IJ}&=-\lim_{\omega^2\to \omega^2_{\pm}}(\omega^2-\omega^2_{\pm})\Delta^{IJ} \\
&=\pm\,\frac{1}{\omega^2_{+}-\omega^2_{-}}\left(\begin{array}{cc}
\omega^2_{\pm}-\beta k^2 & 2\gamma\omega_{\pm} k \\
2\gamma\omega_{\pm} k & \omega^2_{\pm}-\alpha k^2 \end{array}\right)\,.
\eal\eeq
By construction these matrices have zero determinant, meaning that each has a single non-zero eigenvalue and so there are two non-zero norms, as expected. We find these to be
\beq
\lambda_{\pm}=1\pm\,4\gamma^2\left[(\alpha+\beta+4\gamma^2)^2-4\alpha\beta\right]^{-1/2}\,.
\eeq
Clearly the stable-gradients condition ensures that $\lambda_{+}>0$, but the condition $\lambda_{-}>0$ gives an independent constraint,
\beq\bal
\lambda_{-}>0\qquad&\Leftrightarrow\qquad (\alpha+\beta+4\gamma^2)^2>4\alpha\beta+16\gamma^4 \\
&\Leftrightarrow\qquad (\alpha-\beta)^2+8\gamma^2(\alpha+\beta)>0 \,.
\eal\eeq
This is automatically satisfied if $\alpha,\beta>0$, but is otherwise non-trivial. Note also that this inequality implies the second one in \eqref{eq:app grad instability} but is more restrictive than it.

%%%%%%%%%%%%%%%%%%%%%%%%%%%%%%%%%%%%%%%%%%%
%%%%%%%%%%%%%%%%%%%%%%%%%%%%%%%%%%%%%%%%%%%

\section{Analysis of stability conditions} \label{sec:appD}

As explained in the Letter, the stability of the system under consideration hinges on the signs of the functions $\mathcal{H}_1$, $\mathcal{H}_2$, $\mathcal{N}_m$, $\mathcal{N}_j$, $\mathcal{M}_1$, $\mathcal{M}_2$ and $\mathcal{G}_1$ defined in eqs.\ \eqref{eq:H,N defs} and \eqref{eq:G,M defs}. According to our stability criteria, these functions must all be positive definite in the domain of interest, translating into bounds on $G_6$ and $G_{4,X}/\mu^2$. We next provide these bounds for the RN and TOV metrics that we focused on in the Letter.

\subsection{Reissner-Nordstr\"om metric}

For the RN spacetime the domain of interest is $r\geq r_{+}$, with the location of the event horizon being given by $r_{+}=\frac{r_s}{2}\left(1+\sqrt{1-\frac{r_Q^2}{r_s^2}}\right)$. Table \ref{tab:RN stability functions} displays the conditions for the functions determining the stability criteria to be positive definite (recall that $\mathcal{H}_1=\mathcal{H}_2$ and $\mathcal{M}_1=\mathcal{M}_2$ for the RN metric).
\begin{table}[]
\centering
\begin{tabular}{c|c}
\hline
Function        & Condition for positive definiteness \\ \hline\hline
$\mathcal{H}_{1,2}$ & $\frac{G_6}{r_s^2}<\begin{cases}
\frac{\left(1+\sqrt{1-\rho_Q^2}\right)^4}{8\left(1-\rho_Q^2+\sqrt{1-\rho_Q^2}\right)} & \mbox{if $0\leq \rho_Q\leq \frac{\sqrt{15}}{4}$} \\
\frac{32\rho_Q^6}{27} & \mbox{if $\frac{\sqrt{15}}{4} \leq \rho_Q\leq 1$}
\end{cases}$ \\ \hline
\multirow{4}{*}{$\mathcal{N}_j$} 
& $ \frac{G_6}{r_s^2}>-\begin{cases}
\frac{\left(1+\sqrt{1-\rho_Q^2}\right)^4}{8\left(2-3\rho_Q^2+2\sqrt{1-\rho_Q^2}\right)} & \mbox{if $0\leq \rho_Q\leq \frac{\sqrt{3}}{2}$} \\
2\rho_Q^6 & \mbox{if $\frac{\sqrt{3}}{2} \leq \rho_Q\leq \frac{2\sqrt{2}}{3}$}
\end{cases}$ \\ 
& $- 2 \rho_Q^6 <\frac{G_6}{r_s^2}<\frac{\left(1+\sqrt{1-\rho_Q^2}\right)^4}{8\left(3\rho_Q^2-2-2\sqrt{1-\rho_Q^2}\right)}\qquad \mbox{if $\frac{2\sqrt{2}}{3} < \rho_Q\leq 1$}$ \\ \hline
$\mathcal{G}_1$ & $\frac{G_6}{r_s^2}>-\frac{\left(1+\sqrt{1-\rho_Q^2}\right)^4}{8\left(2-\rho_Q^2+2\sqrt{1-\rho_Q^2}\right)}$ \\ \hline
$\mathcal{M}_{1,2}$ & $\frac{G_{4,X}}{\mu^2r_s^2}>-\frac{\left(1+\sqrt{1-\rho_Q^2}\right)^4}{8\rho_Q^2}$ \\ \hline
$\mathcal{N}_m$ & $\frac{G_{4,X}}{\mu^2r_s^2}<\frac{\left(1+\sqrt{1-\rho_Q^2}\right)^4}{8\rho_Q^2}$ \\ \hline
\end{tabular}
\caption{Functions defining the dispersion relations for the RN black hole spacetime and the conditions under which they are positive definite in the domain $r_{+}\leq r <\infty$. Here $\rho_Q\equiv r_Q/r_s$.}
\label{tab:RN stability functions}
\end{table}

\subsection{TOV metric}

The TOV metric components are given by
\beq
f=e^{2\phi}\,,\qquad g=1-\frac{\widetilde{m}}{\widetilde{r}}\,,
\eeq
where here and below all tilde variables correspond to quantities normalized by an arbitrary mass scale $M_{*}$ and associated distance scale $r_{s,*}=2GM_{*}$; this is of course not necessary but is convenient in order to only deal with dimensionless variables in numerical computations.

The functions $\phi$ and $\widetilde{m}$, along with the pressure $\widetilde{p}$, are determined by the TOV equations
\beq\bal
\frac{d\widetilde{m}}{d\widetilde{r}}&=4\pi \widetilde{r}^{\,2}\widetilde{\rho} \,, \\
\frac{d\widetilde{p}}{d\widetilde{r}}&=-(\widetilde{\rho}+\widetilde{p})\frac{\widetilde{m}+4\pi \widetilde{r}^{\,3}\widetilde{p}}{2\widetilde{r}(\widetilde{r}-\widetilde{m})} \,, \\
\frac{d\phi}{d\widetilde{r}}&=-\frac{1}{\widetilde{\rho}+\widetilde{p}}\,\frac{d\widetilde{p}}{d\widetilde{r}} \,.
\eal\eeq
Assuming regularity at the star's center, $\widetilde{r}=0$, one can solve these equations in power series to find
\beq\bal
\widetilde{m}&=\frac{4\pi}{3}\,\widetilde{\rho}_c\widetilde{r}^{\,3}+\cdots \,,\\
\widetilde{p}&=\widetilde{p}_c-\frac{\pi}{3}(\widetilde{\rho}_c+\widetilde{p}_c)(\widetilde{\rho}_c+3\widetilde{p}_c)\widetilde{r}^{\,2}+\cdots \,,\\
\widetilde{\rho}&=\widetilde{\rho}_c-\frac{\pi}{3}\,\widetilde{\rho}^{\,\prime}_c(\widetilde{\rho}_c+\widetilde{p}_c)(\widetilde{\rho}_c+3\widetilde{p}_c)\widetilde{r}^{\,2}+\cdots \,,\\
\eal\eeq
where $\widetilde{\rho}_c:=\widetilde{\rho}(\widetilde{p}_c)$ and $\widetilde{\rho}^{\,\prime}_c:=\frac{d\widetilde{\rho}}{d\widetilde{p}}\Big|_{\widetilde{p}_c}$ are to be obtained from the equation of state $\widetilde{\rho}(\widetilde{p})$. It is then straightforward to evaluate the stability criteria at $\widetilde{r}=0$ in order to derive the bounds quoted in the main text.

%%%%%%%%%%%%%%%%%%%%%%%%%%%%%%%%%%%%%%%%%%%
%%%%%%%%%%%%%%%%%%%%%%%%%%%%%%%%%%%%%%%%%%%

\bibliographystyle{apsrev4-1}
\bibliography{GPstabilityBiblio}

\end{document}